\documentclass[epj]{svjour}
\usepackage{graphics}
\usepackage{amsmath}
\usepackage{amssymb}
\usepackage{graphicx}
\usepackage{dsfont}
\usepackage{float}
\usepackage{subfig}
\usepackage{placeins}
\usepackage{setspace}
\usepackage{listings}
\usepackage{color}
\usepackage{fancyvrb}
\usepackage{placeins}
\usepackage{pdfpages}
\usepackage{fullpage}
\usepackage{multirow}

\begin{document}
\title{Double Exchange model for nanoscopic clusters}
%\subtitle{Do you have a subtitle?\\ If so, write it here}
\author{D.~Rotter\inst{1}, A.~Valli\inst{1}, G.~Sangiovanni\inst{1}$^,$\inst{2} \and K.~Held\inst{1}% etc
% \thanks is optional - remove next line if not needed
\thanks{\emph{Present address:} held$@$ifp.tuwien.ac.at}%
}                     % Do not remove
\offprints{}          % Insert a name or remove this line
\institute{Institute of Solid State Physics, Vienna University of Technology, 1040 Vienna, Austria 
\and Institute for Theoretical Physics and Astrophysics, University of W\"urzburg, Am Hubland, D-97074 W\"urzburg, Germany}
\date{Received: date / Revised version: date}
% The correct dates will be entered by Springer
%
\abstract{
We solve the double exchange model on nanoscopic clusters exactly, and specifically consider a six-site benzene-like nanocluster.
This simple model is an ideal testbed for studying magnetism in nanoclusters 
and for validating approximations such as the dynamical mean field theory (DMFT).
Non-local correlations arise between neighboring localized spins due to the Hund's rule coupling, 
favoring a short-range magnetic order of ferro- or antiferromagnetic type. 
For a geometry with more neighboring sites 
or a sufficiently strong hybridization between leads and the nanocluster,
these non-local correlations are less relevant, and DMFT can be applied reliably. 
\PACS{
      {PACS-key}{describing text of that key}   \and
      {PACS-key}{describing text of that key}
     } % end of PACS codes
} %end of abstract
\maketitle
\section{Introduction}
\label{sec:intro}
In recent years, manganites such as La$_{1-x}$Ca$_{x}$MnO$_3$ (LCMO) have attracted  great interest, 
mainly due to the colossal magnetoresistance \cite{CMR}. 
In these materials, the crystal field splits the five 3d orbitals of the Manganese atoms 
into two $e_g$ and three $t_{2g}$ orbitals due to the perovskite structure. 
The latter are localized and as a consequence of Hund's exchange half-filled, forming a spin 3/2. 
This spin, the itinerant $e_g$ electrons and their coupling, again by Hund's exchange, 
constitute the double exchange or ferromagnetic Kondo lattice model \cite{DE}. 
This arguably simplest model for manganites gives rise, in the bulk, to a ferromagnetic double exchange since 
a hopping energy of the $e_g$ electrons can be gained only for a ferromagnetic (FM) alignment of the 
$t_{2g}$ spins. For half-filled $e_g$ bands one the other hand, the alignment of the $t_{2g}$ spins is antiferromagnetic (AF) due to superexchange. 

More recently, nanoclusters of manganites have been synthesized; and remarkably, a size-control of the 
charge and magnetic ordering in half-doped LCMO %Ca$_{0.5}$La$_{0.5}$MnO$_3$ 
has been demonstrated \cite{sarkar123104,Das}. 
Hence, size can be utilized to optimize the magnetic properties 
and the magnetoresistance of manganites for technological applications. 
		
Not only these experiments, but also generally the emergence and peculiarity of magnetism in nanoclusters, 
motivates us to study the double exchange model: 
\begin{equation}\label{HFKLM}
 H=-t\sum _{\langle ij\rangle\sigma} c^{\dagger}_{i\sigma}c_{j\sigma} -2J\sum _i s _i S_i .
\end{equation}	
Here, $c^{\dagger}_{i\sigma}$ ($c_{i\sigma}$) are the creation (annihilation) operators 
of an electron at site i with spin $\sigma $, 
and $t$ is the effective hopping amplitude corresponding to the double exchange process. 
$S$ and $s$ are the spins of the localized and mobile electrons respectively, 
$J$ is the Hund's coupling between those spins. 
Henceforth we restrict ourselves to a Ising symmetry of the localized spins so that $S_i$ simplifies to 
$S_i=\pm1 $, and $s_{i}=\frac{1}{2}\sum _{\sigma=\pm1} c^{\dagger}_{i\sigma} \sigma c_{i\sigma}$. Note the results only depend on $|S|J$ so that taking $S_i=\pm1 $ instead of $S_i=\pm 3/2$ only corresponds to a redefinition of $J$.

\begin{figure}
 \begin{center}
  \includegraphics[width=0.5\textwidth]{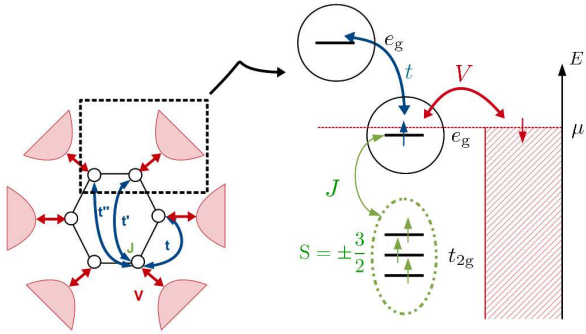}
  \caption{Schematic representation of the model system consisting of six sites, 
each of them is connected to a lead via the hybridization strength $V_{i\eta k}=V\delta_{i\eta}$, 
and electrons can hop between the sites $i$ and $j$ of the nanostructure via hopping channels $t_{ij}$. 
Itinerant electrons interact with localized spins $S$ at each site via Hund's coupling $J$.\label{MyKondo}}
 \vspace{-17pt}
 \end{center}
\end{figure}

Besides these usual terms of the double exchange model, we consider 
a hybridization $V_{i\eta k}$ connecting each site $i$ of the nanostructure 
to some non-interacting environment $\eta$ with eigenenergies $\epsilon_{\eta k       }$ where $k$ labels the different states in the respective environment,
see Fig. \ref{MyKondo}. Such term allows one to describe, e.g., a surface 
or electrodes (leads) applied to the system for electric transport measurements. 
Altogether, this leads to the Hamiltonian 
\begin{equation}\label{HKondo}
\begin{split}
 H &= \sum _{ij \sigma} t_{ij}c^{\dagger}_{i\sigma}c^{\phantom{\dagger}}_{j\sigma} 
    - 2 J\sum_i s_i S_i \\
   &+ \sum_{i\eta k\sigma} V_{i\eta k} c^{\phantom{\dagger}}_{i\sigma} a^{\dagger}_{\eta k\sigma} + H.c. 
    + \sum_{\eta k\sigma} \epsilon_{\eta k} a^{\dagger}_{\eta k\sigma}a^{\phantom{\dagger}}_{\eta k\sigma} .
\end{split}
\end{equation}
In analogy to Refs. \cite{nanoDGA} we consider a nanoscopic system 
made of six-sites arranged in a ring structure. 
For the sake of simplicity, we will restrict ourselves to a configuration in which all sites are equivalent 
and each site is connected to its own paramagnetic metallic lead, i.e., $V_{i\eta k}\!=\!V\delta_{i\eta}$ , and take a constant density of states 
 $\rho=1/(2D)$ for the non-interacting bath, where $D$ is the bath half-bandwidth. 
%We also do not include the Jahn-Teller coupling and the Coulomb interaction between $e_g$ electrons, 
%both of which have been shown to be important for a more complete description of manganites \cite{MillisJT,Held}, 
%but focus on the physics emerging from $J$ only. 
As we will see, the competition between kinetic energy, hybridization, and Hund's exchange exhibits  
some very interesting properties, and can lead, depending on the parameters, 
to ferromagnetic (FM) or antiferromagnetic (AF) (and of course paramagnetic) short-range ordering of the localized spins. 

In this paper, we solve the model (\ref{HKondo}) exactly on a six-site ring, 
and take it as a benchmark for testing the validity of a dynamical mean field theory (DMFT) \cite{DMFT} 
approximation. Let us note that the DMFT solution for the Kondo lattice model in the bulk 
was derived before by Furukawa \cite{Furukawa1994}. In order to apply DMFT to a finite system, 
we employ its recently introduced nanoscopic version \cite{nanoDGA}, 
which has been shown to be suitable to deal with  complex nanostructures. 
The outline of the paper is as follows: 
In Sec. \ref{sec:method}  we present the path integral method for the exact solution, and the DMFT scheme. 
In Sec. \ref{sec:results:exact} we present the results obtained solving the model exactly, 
and we use them as a benchmark for DMFT in Sec. \ref{sec:results:DMFT}. 
Finally, we summarize our results in Sec. \ref{sec:conclusion}.

\section{Method}
\label{sec:method}
Let us start with some simple symmetry considerations. 
In a six-sites system, with two possible alignments for each localized spin $S_i$, 
there are $2^6=64$ different possible configurations $\{S_i\}$. 
Since in the (high temperature) paramagnetic phase the $Z_2$ rotational symmetry is conserved, 
the overall alignment of the spins does not matter, thereby reducing the number of inequivalent 
configurations by a factor of two. 
In the following, 0 and 1 will represent the two different orientations of the localized spins. 
So, e.g., a configuration with only one spin-up would be represented by \{100000\}, which, 
in the paramagnetic phase, is also equivalent to \{011111\}.
% Moreover, if the system displays translational symmetry, the exact position of the spins is not important, 
%so that \{100000\} and \{010000\} are degenerate. Moreover, in a first approximation, configurations 
%with the same number of ferro- (FM) and antiferromagnetic (AF) bonds are equivalent.

In order to check the role of the connectivity, 
in the following we will consider similar cases  as in Ref. \cite{nanoDGA}: 
(i) hopping is only possible to next neighbors ("NN~t") 
in which the number of inequivalent spin configuration is further reduced to eight, by symmetry, and 
(ii) hopping is equally possible to all sites ("all~t"), 
with only four inequivalent spin configurations. 
The latter hopping topology is not a realistic one, as the hopping amplitude is a decreasing function 
of the inter-site distance. 
However it will be useful to study, in the philosophy of Ref. \cite{nanoDGA}, the effect of enhanced connectivity 
without changing the geometry of the nanostructure. 
All these symmetry considerations are summarized in Table \ref{Tab:1}. 
%\begin{widetext}
 \begin{table*}
  \centering
  \begin{tabular}{c c c c c}
   \cline{2-5}
   \multicolumn{1}{c}{} & \multicolumn{2}{|c|}{NN~t} & \multicolumn{2}{|c|}{all~t} \\ \cline{2-5} \hline
   \multicolumn{1}{|c|}{configuration} & \multicolumn{1}{|c|}{degeneracy} & \multicolumn{1}{|c|}{$\sharp$ of AF bonds} & \multicolumn{1}{|c|}{degeneracy} & \multicolumn{1}{|c|}{$\sharp$ of AF bonds} \\ \hline
   \multicolumn{1}{|c|}{\{000000\} }& \multicolumn{1}{|c|}{2} & \multicolumn{1}{|c|}{0/6} & \multicolumn{1}{|c|}{2} & \multicolumn{1}{|c|}{0/15}\\ \hline
   \multicolumn{1}{|c|}{\{100000\}} &\multicolumn{1}{|c|} {12} &\multicolumn{1}{|c|}{2/6} & \multicolumn{1}{|c|}{12} &\multicolumn{1}{|c|}{5/15}\\ \hline
   \multicolumn{1}{|c|}{\{110000\} }& \multicolumn{1}{|c|}{12 }& \multicolumn{1}{|c|}{2/6} &\multicolumn{1}{|c|}{ 30} &\multicolumn{1}{|c|}{ 8/15}\\ \hline
   \multicolumn{1}{|c|}{\{111000\} }& \multicolumn{1}{|c|}{6 }&\multicolumn{1}{|c|}{2/6} &\multicolumn{1}{|c|}{ 20} &\multicolumn{1}{|c|}{ 9/15}\\ \hline
   \multicolumn{1}{|c|}{\{101000\} }& \multicolumn{1}{|c|}{12 }& \multicolumn{1}{|c|}{4/6} & \multicolumn{2}{c}{ }\\ \cline{1-3}
   \multicolumn{1}{|c|}{\{100100\} }& \multicolumn{1}{|c|}{6 }& \multicolumn{1}{|c|}{4/6} &\multicolumn{2}{c}{ }\\ \cline{1-3}
   \multicolumn{1}{|c|}{\{101100\} }& \multicolumn{1}{|c|}{12 }& \multicolumn{1}{|c|}{4/6} &\multicolumn{2}{c}{ }\\ \cline{1-3}
   \multicolumn{1}{|c|}{\{101010\} }&\multicolumn{1}{|c|}{ 2} & \multicolumn{1}{|c|}{6/6} &\multicolumn{2}{c}{ }\\ \cline{1-3}
  \end{tabular}
 \caption{Non-equivalent localized spin configurations for the six-site ring for both NN~t and all~t hopping topologies, 
 taking into account $Z_2$ rotational symmetry and translational invariance.\label{Tab:1}}
\end{table*}
%\end{widetext}

For an isolated nanostructure, i.e., without hybridization to the leads, 
the double exchange model can be solved exactly by calculating all eigenvalues and eigenstates of the Hamiltonian (\ref{HKondo}). 
This procedure becomes however computationally unfeasible if the number of sites of the nanostructure is very large, 
or if non-interacting leads are included, as it would require many additional bath-sites 
for an accurate description of the leads. We therefore employ a Green's function technique, 
were the non-interacting degrees of freedom can be integrated out exactly. 
In the Grassmann path-integral formulation \cite{NegOrl} the expectation value of an observable ${\cal O}$ 
is calculated as
\begin{align}\label{observable}
 \langle{\cal O}\rangle = & \frac{1}{Z} \displaystyle \sum_{\{S_i\}} \int {\cal O} e^{-\mathcal{S}^{\{S_i\}}} ,
\end{align}
where $Z=\sum_{\{S_i\}}\int e^{-\mathcal{S}^{\{S_i\}}}$ denotes the partition function of the system 
and the integration symbol 
$\int \equiv 
 \prod_{i \sigma}      \int \mathcal{D}[c^{\dagger}_{i \sigma}, c^{\phantom{\dagger}}_{i \sigma}] 
 \prod_{\eta k \sigma} \int \mathcal{D}[a^{\dagger}_{\eta k \sigma}, a^{\phantom{\dagger}}_{\eta k \sigma}]$ 
represents the functional integral extending over the Grassmann variables 
$c^{\dagger}_{i \sigma}$ ($c^{\phantom{\dagger}}_{i \sigma}$) and 
$a^{\dagger}_{\eta k \sigma}$ ($a^{\phantom{\dagger}}_{\eta k \sigma}$) associated to the
fermionic  creation (annihilation) operators. 
Finally, the action is given by
\begin{align}\label{action}
 %\mathcal{S}(\{ c^{\dagger},c,a^{\dagger},a\} ) 
 \mathcal{S}^{\{S_i\}}\!=\!\int\limits_0^\beta \! d\tau  
   & \Big[ \sum_{i\sigma }c^{\dagger}_{i\sigma}(\tau)(\partial_\tau-\mu)c^{\phantom{\dagger}}_{i\sigma}(\tau) \nonumber \\
 + & \sum_{i\neq j,\sigma } c^{\dagger}_{i\sigma}(\tau) t_{ij} c^{\phantom{\dagger}}_{j\sigma}(\tau) 
  -2 \sum_{i} J s_{i}(\tau) S_i \nonumber \\
 + & \sum_{\eta k \sigma }a^{\dagger}_{\eta k \sigma}(\tau)(\partial_\tau + \epsilon_{\eta k} - \mu )a^{\phantom{\dagger}}_{\eta k \sigma}(\tau) \nonumber \\
 + & \sum_{i \eta k \sigma }V_{i \eta k}(c^{\dagger}_{i\sigma}(\tau)a^{\phantom{\dagger}}_{\eta k \sigma}(\tau)+h.c.) \Big] ,
\end{align}
where $\tau\!\in\![0,\beta)$ is the imaginary time, $\beta=1/T$ is the inverse temperature, 
and $\mu$ is the equilibrium chemical potential. 
Hence the Green's function is defined as 
\begin{align}\label{eq:GFdef}
 G_{m n \sigma}(\tau) = & \frac{1}{Z} 
  \sum_{\{S_i\}} \int c^{\phantom{\dagger}}_{m\sigma}(\tau) c^{\dagger}_{n\sigma}(0) e^{-\mathcal{S}^{\{S_i\}}} .
\end{align}
%the Green's function can be written as 
% \begin{align}\label{Greensfunction}
% G_{\mu \nu \sigma}(\tau) = & \frac{1}{Z} \displaystyle \sum_{\{S_i\}}\int \displaystyle\prod_{i k \sigma} 
%   \mathcal{D}[c^{\dagger}_{i\sigma}, c^{\phantom{\dagger}}_{i\sigma}] \mathcal{D}[a^{\dagger}_{ik\sigma}, a^{\phantom{\dagger}}_{ik\sigma}] \nonumber \\
%   & \times c^{\phantom{\dagger}}_{\mu\sigma}(\tau) c^{\dagger}_{\nu\sigma}(0) e^{-\mathcal{S}^{\{S_i\}}} ,
% \end{align}
%where  $c^{\dagger}_{i\sigma}$, $a^{\dagger}_{ik\sigma}$ ($c^{\phantom{\dagger}}_{i\sigma}$,  $a^{\phantom{\dagger}}_{ik\sigma}$) now denotes a 
%Grassmann variable which relates to the corresponding creation (annihilation) operator, Z the partition function of the system
%\begin{equation}\label{partfunc}
% Z=\int \displaystyle\prod_{ik\sigma} \mathcal{D}[c^{\dagger}_{i\sigma}, c^{\phantom{\dagger}}_{i\sigma}] 
%        \mathcal{D}[a^{\dagger}_{ik\sigma}, a^{\phantom{\dagger}}_{ik\sigma}] 
%        \displaystyle\sum_{\{S_i\}} e^{-\mathcal{S}^{\{S_i\}}},
%\end{equation}
%and the action is given by

\noindent Since the non-interacting electron operators only enter quadratically in the action \eqref{action}, 
we can integrate them out by a simple Gaussian integral \cite{NegOrl}, yielding the effective action 
\begin{align}\label{effaction}
 \mathcal{S}_{\textrm{eff}}^{\{S_i\}} 
  &\!=\!\int_0^\beta \!d\tau \Big[ 
    \sum_{i\sigma}c^{\dagger}_{i\sigma}(\tau)(\partial_\tau-\mu)c^{\phantom{\dagger}}_{i\sigma}(\tau) \nonumber \\
  & + \sum_{ij\sigma} c^{\dagger}_{i\sigma}(\tau) t_{ij} c_{j\sigma}(\tau)
    - 2J \sum_{i} S_i s_i(\tau) \big) \Big] \nonumber \\
  & + \int_0^\beta \!d\tau \! \int_0^\beta \!d\tau^\prime 
      \sum_{ij \sigma} c^{\dagger}_{i\sigma}(\tau) \Delta_{ij}(\tau-\tau\prime) 
                      c^{\phantom{\dagger}}_{j\sigma}(\tau^\prime) .
\end{align}
Here the hybridization function $\Delta_{ij}(\tau)$ takes into account the virtual processes 
of itinerant electrons hopping back and forth between the leads and the nanostructure.
Its Fourier transform, analytically continued to the real axis, corresponds to the retarded function
\begin{equation}
 \Delta_{ij}(\omega)=\sum_{\eta k}
   \frac{V^{\phantom{\dagger}}_{i \eta k} V^*_{j \eta k}}{\omega+\imath\delta-\epsilon_{\eta k}}.
\end{equation}
At this point it is convenient to recast the Green's function of the nanostructure \eqref{eq:GFdef}  
in terms of the effective action \eqref{effaction} considering the contribution 
of each configuration $\{S_i\}$ of the localized spins explicitly
\begin{eqnarray}\label{eq:GF}
 G_{mn \sigma}(\tau) = \sum_{\{S_i\}}P^{\{S_i\}} G_{mn \sigma}^{\{S_i\}}(\tau),
\end{eqnarray}
via a functional integral for each configuration ${\{S_i\}}$
\begin{equation}\label{eq:GFconfig}
 G_{mn \sigma}^{\{S_i\}}(\tau) = \frac{1}{Z^{\{S_i\}}} 
    \int c^{\phantom{\dagger}}_{m\sigma}(\tau ) c^{\dagger}_{n\sigma}(0) e^{-\mathcal{S}_{\textrm{eff}}^{\{S_i\}}}
\end{equation} 
where we now only integrate  the interacting degrees of freedom, i.e.,
$\int \equiv \prod_{i \sigma}  \int  \mathcal{D}[c^{\dagger}_{i \sigma}, c^{\phantom{\dagger}}_{i \sigma}]$. 
The weights of the configuration-dependent Green's functions in Eq. \eqref{eq:GF} are defined by
\begin{equation}\label{GProb}
 P^{\{S_i\}}=\frac {Z^{\{S_i\}}}{\displaystyle\sum_{\{S_i\}}Z^{\{S_i\}}}, 
\end{equation} 
where
\begin{equation}\label{partfunceff}
 Z=\sum_{\{S_i\}} Z^{\{S_i\}} = \sum_{\{S_i\}} \int e^{-\mathcal{S}_{\textrm{eff}}^{\{S_i\}}}.
\end{equation}
The calculation of the probabilities Eq.\ (\ref{GProb}) requires the evaluation 
of the determinant of $S_{\textrm{eff}}^{\{S_i\}}$. This is more conveniently done in  Fourier space 
%of (Matsubara) frequencies $\omega_n\!=\!(2n+1)\frac{\pi}{\beta}$, with $\beta=1/T$ is the inverse temperature. 
where it takes the matrix form, in site indices $i$ and $j$, 
\begin{equation}
 \mathcal{S}_{ij\sigma}(\omega)=(-\omega-\mu + \Delta_{ij \sigma}(\omega)-J\sigma S_i)\delta_{ij}+t_{ij} .
\end{equation}	
Calculating these determinants (probabilities) as well as the Green's function 
for the inequivalent of the $2^6$ spin configurations
$\{S_i\}$ yields the exact summation of the Green's function of the nanocluster coupled to non-interacting leads.

Besides the exact solution, we have also employed DMFT. 
There are several versions of DMFT for treating nanoscopic or spatially 
inhomogeneous systems \cite{nanoDGA,Florens08a,Potthoff99,Snoek08} with minor differences. 
Here, we perform our calculations in the same fashion as in Ref. \cite{nanoDGA}, 
restricting ourselves to the 1-particle, i.e. DMFT, realization of 
the dynamical vertex approximation \cite{toschiPRB75,kataninPRB80} 
applied to nanoscopic systems.
%Here, for each site $i$ a non-interacting 
%impurity Green function ${\mathcal{G}^0_i}^{-1}$ is determined, from this 
%the interacting Green function is determined at frequency $\omega$ for the 
%double exchange model as 
The starting point of the DMFT scheme is the Green's function of the nanostructure, 
of which the generic matrix element of its inverse reads
\begin{equation} \label{Dyson}
 \displaystyle{ \big\{{G}^{-1}\big\}_{ij\sigma}(\omega) = 
   \omega \delta_{ij}-t_{ij}-\Delta_{ij \sigma}(\omega)-\Sigma_{ij\sigma}(\omega)} ,
\end{equation}
where $\Sigma$ denotes the self-energy. 
For each of the (inequivalent) sites of the nanostructure, we define a local problem 
determining the associated Weiss field ${{\cal G}_i^0}^{-1}\!=\!{[G_{ii}]}^{-1}\!+\!\Sigma_{ii}$ 
out of the local block of site $i$ of the Green's function $G_{ii}$, 
with an initial guess of the (local) self-energy $\Sigma_{ii}$ (usually zero). 
This procedure maps each atom onto an effective one-body local problem 
where the local spin can either point up or down.
Let us note that the DMFT approximation includes the local spins. That is,
DMFT substitutes the actual  localized spin configuration on the other sites 
by an effective Weiss field ${\mathcal{G}}^0$, which in the paramagnetic phase is spin-independent. Hence localized spin up and down have the same probability 
and the interacting DMFT  Green function of site $i$ reads  \cite{Furukawa1994}
\begin{equation}\label{ImpSolver}
 G_i(\omega)=\frac{1/2}{{\mathcal{G}^0_i}^{-1}(\omega)+J}+\frac{1/2}{{\mathcal{G}^0_i}^{-1}(\omega)-J}.
\end{equation}
Since this has the form of a non-interacting Green's function averaged over a potential $\pm J$, Eq. (\ref{ImpSolver}) is temperature independent 
(which does not hold for DMFT Green's function in general). 
Mathematically this is equivalent to binary disorder. Hence, 
Eq.\ (\ref{ImpSolver}) allows us to determine a local self-energy 
$\Sigma_{ii}\!=\!{{\cal G}^0_i}^{-1}+{G_{i}}^{-1}$ for each site $i$. 
Using the calculated self-energies as an input for the Dyson equation \eqref{Dyson}
we can finally calculate a new Green's function for the nanostructure, 
and iterate the cycle self-consistently until convergence. \cite{nanoDGA} 
In the case we are considering
all sites are equivalent due to the symmetry of the problem. Therefore we need to solve only one local problem, 
yielding a local self-energy which is the same for each site of the nanostructure.

\subsection{Results: Exact Solution} \label{sec:results:exact}
\begin{figure*}[ht!]
 \begin{center}		 
 \includegraphics[angle=270,width=1.0\textwidth]{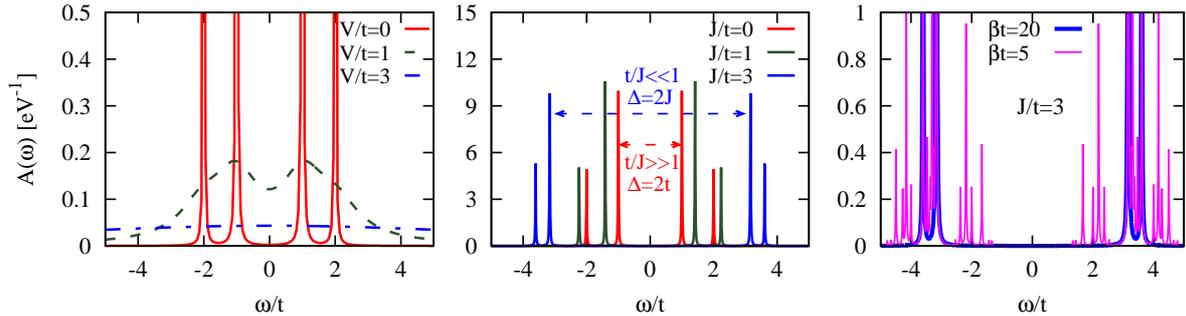}
 \caption{Spectral function $A(\omega )$ for NN~t in the broad-band limit $D\!=\!20t$. 
Left panel: $J\!=\!0$;  the non-interacting levels are broadened 
either by the hybridization $V$ or by a tiny cut-off $\delta/t\!=\!0.01$. 
%Increasing $V/t$ the system forms broad bands and becomes a featureless spectrum in the limit $V\!\gg\!t$. 
%In absence of interaction the model is independent of temperature. 
Middle panel: $V\!=\!0$ for different values of $J$ and $\beta t\!=\!20$. 
One can clearly see that the gap $\Delta$ is controlled by $t$ or $J$ in the weak and strong coupling regimes, respectively. 
Right panel: Temperature dependence of the spectral function at $J\!=\!3t$ and $V\!=\!0$.}
.\label{fig:spectraES}
 \end{center}
\end{figure*}

In this section the exact results obtained are presented. 
As already mentioned, there are several competing energy scales in the problem. 
Therefore it is useful to describe the non-interacting ($J/t=0$) and isolated ($V/t=0$) ``benzene'' molecule first, 
and analyze separately the effects of the hybridization and of the interaction. 
The following analysis is summarized in the spectral function, connected to the retarded Green's function 
via the relation $A(\omega)=-\frac{1}{\pi}G_{ii}(\omega+\imath\delta)$, shown in Fig. \ref{fig:spectraES}.

When only the kinetic term of Hamiltonian \eqref{HKondo} is taken into account, 
the nanostructure is translational invariant and the wavevectors $k\!=\!m\pi/3$ ($m=0,\ldots,5$) 
are conserved quantum numbers. Setting the chemical potential $\mu\!=\!0$, yields the dispersion relation 
\begin{equation}\label{EkFT}
 E(k)=-2t\cos (k)-2t'\cos (2k)-t''\cos (3k) ,
\end{equation}
where, $t$, $t'$, and $t''$ are the nearest neighbor, the next-nearest neighbor, 
and the next-next-nearest neighbor hopping amplitudes, respectively. 
In the following $t$ sets our unit of energy, and the hopping configurations as defined by: 
(i) NN~t case: $t'=t''=0$, and (ii) all~t case: $t'=t''=t$. 
The spectral function associated to this system is constituted by $\delta$-like peaks 
corresponding to the six energy eigenstates (some of which are degenerate), 
and it is shown in the left panel of Fig. \ref{fig:spectraES} for the NN~t case (solid line).
It is important to note that in both the NN~t and all~t case the spectral weight at the Fermi energy ($E_F$) 
is vanishing and the system is a band insulator. 

In order to show what happens when the structure is connected to metallic leads, 
we restrict ourselves to the NN~t case, but analogous considerations apply to any hopping topology. 
As already mentioned, the leads are bulk systems of non-interacting electrons 
described by a flat density of states $\rho\!=\!1/(2D)$. 
We consider the broad-band limit $D\!\gg\!t$ (in order to avoid the particular physics 
arising at the edge of the leads' conduction band) and we set $D\!=\!20t$ in the following.  
The most important effect of the hybridization between the ring and the non-interacting environment 
is the broadening of all peaks in the spectral function 
which are within the bandwidth of the leads, while the shift of the eigenenergies, 
i.e. the real part of the hybridization function, is $O(\omega/D)$. 
With increasing $V$, the discrete spectrum of the molecule 
evolves into a broad band (see Fig. \ref{fig:spectraES}, left panel).

The next step is to study what is the effect of the Hund's coupling $J$ on such a nanoscopic system. 
In general, for a system of N sites, there are $2^N$ different configurations of the localized spins, 
which means that there are $2N2^N$ eigenstates. 
Most of them are degenerate in the non-interacting, paramagnetic system with translational invariance. 
In the presence of $J$ and for a given configuration  $\{S_i\}$, translational invariance is broken,
and some of those degeneracies are lifted. 
At low enough temperatures, however, only the spin configurations corresponding to the lowest energy 
will be populated, so that the main effect of $J$ is to increase the size of the gap. 
For large $J$, one can show that the size of the gap is $2J$ (see Fig. \ref{fig:spectraES}, middle panel). 
At higher temperature ($\beta t\!=\!5$), due to the broadening of the Fermi-Dirac distribution, 
more states are occupied, and more energy eigenstates are visible 
in the spectral function (see Fig. \ref{fig:spectraES}, right panel).
 
Moreover, due to the Hund's exchange that couples locally the itinerant and localized spins degrees of freedom, 
non-local (short-range) magnetic correlations arise between the localized spins, 
so that they mutually influence their orientation. 
We will see that, depending on the ratio between $J$ and $V$, and on the density of the itinerant electrons, 
those correlations can favor a FM or AF alignment of the localized spins. 
The exact summation technique described in Sec. \ref{sec:method} allows us to calculate 
the probabilities $P[\{S_i\}]$ of Eq.\eqref{GProb}, 
in order to show which spin configurations are energetically more favorable. 
As the probability of a configuration is -to a good approximation- only depending on the number of FM and AF bonds, 
regardless of the exact relative position of these bonds, 
we can identify four representative spin configurations, for each of the hopping topologies introduced above.

In the following we discuss general consideration in order to understand (or predict) 
what kind of magnetic correlation will affect the localized spins. 
In the isolated system (i.e. $V/t=0$) only two energy scales are competing, namely the hopping $t$ which tends 
to minimize the energy and delocalizes the itinerant electrons, and the Hund's coupling $J$, which couples their spins 
with the localized ones. 
Due to the Pauli principle, hopping processes between neighboring sites are only possible if the sites are 
occupied with one electron or two electrons with opposite spins. 
This leads to an effective AF correlation between the itinerant electrons, 
and an AF alignment is particularly favored at half-filling, when there is on average one electron per site. 
At the same time, in order to further minimize the energy, each localized spin tends to align to the spin 
of the itinerant electron (density) at the corresponding site. 
This mechanism leads to AF (superexchange) correlation between neighboring localized spins. 
On the other hand, when the system is out of half-filling, the presence of empty (or doubly occupied) sites 
favors instead a FM double exchange correlation between localized spins. 
According to the considerations above, it is therefore interesting to study in detail the NN~t case at 
non-integer density, e.g., quarter-filling, as well.
The role of the hybridization is to effectively decouple the itinerant electron at each site
from the other sites and the localized spin. The hybridization hence suppresses non-local magnetic correlations and
leads to a more evenly distribution of  spin configuration probabilities $P[\{S_i\}]$.
\begin{figure}
 \begin{center}
 \includegraphics[angle=270,width=0.5\textwidth]{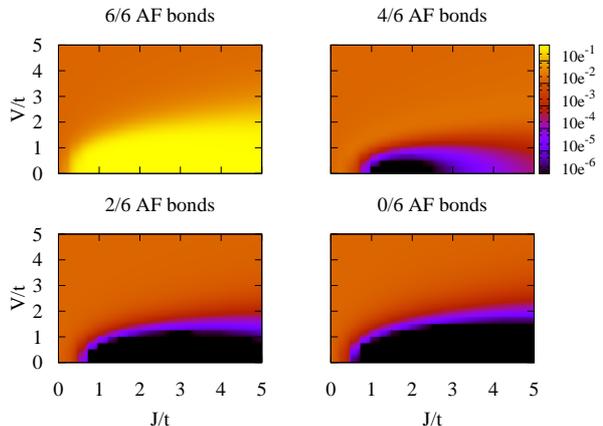}
 \caption{Probability $P[\{S_i\}]$ for four representative configurations (out of $2^6$) of localized spins, 
classified by the number of AF bonds, as a function of $J$ and $V$, at $\beta t\!=\!20$ and half-filling, in the NN~t hopping topology. 
An AF kind of correlation of the localized spins is preferred for $J \gtrsim t$ and $J \gtrsim V$.}
 \label{BBPhaseb20nn}
 \end{center}
\end{figure}
\begin{figure}
 \begin{center}
 \includegraphics[angle=270,width=0.5\textwidth]{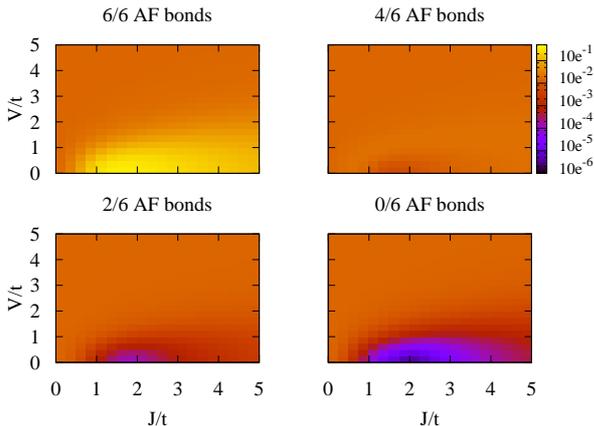}
 \caption{As in Fig. \ref{BBPhaseb20nn} but at $\beta t\!=\!5$. 
At high temperatures the system is characterized by more evenly distributed probabilities 
for the configurations of the localized spins.}
 \label{BBPhaseb5nn}
 \end{center}
\end{figure}
In the following we analyze the probabilities of the different localized spins configurations 
as a function of V and J for NN~t  both at half- and quarter-filling. 
In Fig. \ref{BBPhaseb20nn} we show that for NN~t at half-filling 
and $\beta t=20$, an AF (short range) ordering of the localized spins is energetically favored in a region where $J\gtrsim V$, 
i.e. in the region where the probability of the configuration with six out of six bonds arranged AF is the highest 
(yellow region of the top left panel of Fig. \ref{BBPhaseb20nn}).
When the temperature is raised to $\beta t=5$, states which are higher in energy become populated as well, 
leading to more evenly distributed probabilities, and therefore the AF correlation is weaker or confined to a smaller region 
of the $V\!-\!J$ plane, as shown in Fig. \ref{BBPhaseb5nn}. 
However, one also expects that, independently of temperature, the probabilities of the different configurations 
to become  evenly distributed even in the strong coupling limit $J\gg t$. 
This can be understood by considering that the short-range magnetic order arises due to the \emph{interplay} of $J$ and $t$. 
At $J\gg t$, the energy gain associated to hopping processes can be estimated 
in second order perturbation theory as $\Delta E\approx -t^2/2J$, and is vanishing for $J/t \to \infty$. 
As a consequence, the localized spins become uncorrelated in this limit and non-local correlations are suppressed. 

\begin{figure}
 \begin{center}
 \includegraphics[angle=270,width=0.5\textwidth]{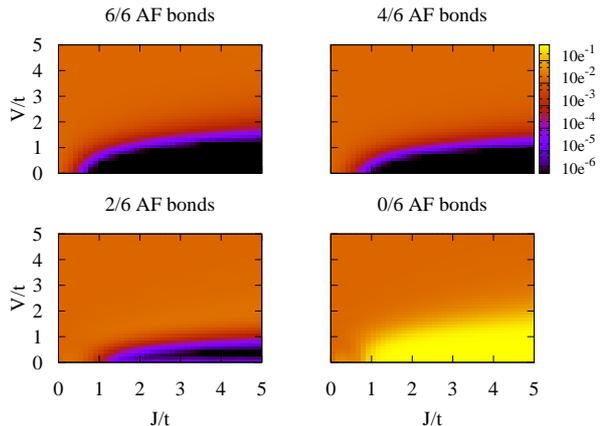}
 \caption{Same as Fig. \ref{BBPhaseb20nn}, but at quarter-filling, where FM (short range) ordering is favored 
when $J$ becomes the dominating energy scale.}
 \label{SBQFPhaseb20nn}
 \end{center}
\end{figure}

In the quarter-filled case we find a similar behavior as for the half-filling, 
but, as expected, with dominant FM rather than AF configuration, 
i.e. the highest probability corresponds to the configuration with zero AF bonds (bottom right panel of Fig. \ref{SBQFPhaseb20nn}).
This result can be understood by considering that, at low densities, the itinerant
electrons have an enhanced  (with respect to half-filling) probability to hop to empty sites.   
These kind of processes are favored if the neighboring localized spins are aligned FM, 
and suppressed if they are aligned AF, since the process would cost an interaction energy of order $2J$. 
Recently the double exchange model for a finite system for fully saturated $t_{2g}$ spins has been analyzed in Ref. \cite{henningPRB86}.

\begin{figure}
 \begin{center}
 \includegraphics[angle=270,width=0.5\textwidth]{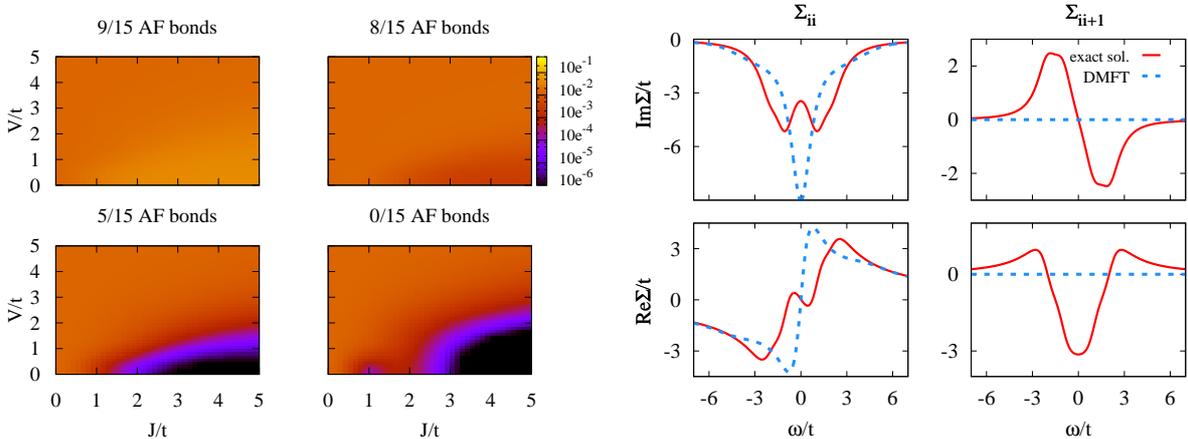}
 \caption{Same as Fig. \ref{BBPhaseb20nn} but in the all~t hopping topology. Increasing the connectivity of the system 
leads to much more evenly distributed probabilities.\label{BBATPhaseb20nn}}
 \end{center}
\end{figure}

As we aim to compare the exact solution to DMFT, we also notice that 
DMFT has the property of becoming exact in the limit of infinite connectivity $z\rightarrow\infty $ \cite{DMFT}. 
One expects therefore that, increasing the connectivity of the system, non-local correlations 
to be averaged out and the mean field approximation to become reliable. 
As already mentioned, a possibility to do this, without changing the geometry of the nanostructure, 
is to introduce longer range hopping. It is therefore interesting to study how the system changes 
when we go from NN~t to all~t, i.e., increasing the connectivity from $z=2$ to $z=5$. 
In Fig. \ref{BBATPhaseb20nn} we show that indeed the probabilities of the different spin configurations 
are in general more evenly distributed than for the NN~t case, even at low temperatures,  
while AF short range order is still slightly favored in the strong coupling limit. 

%\FloatBarrier

\begin{figure*}
 \begin{center}
 \includegraphics[angle=270,width=0.3\textwidth]{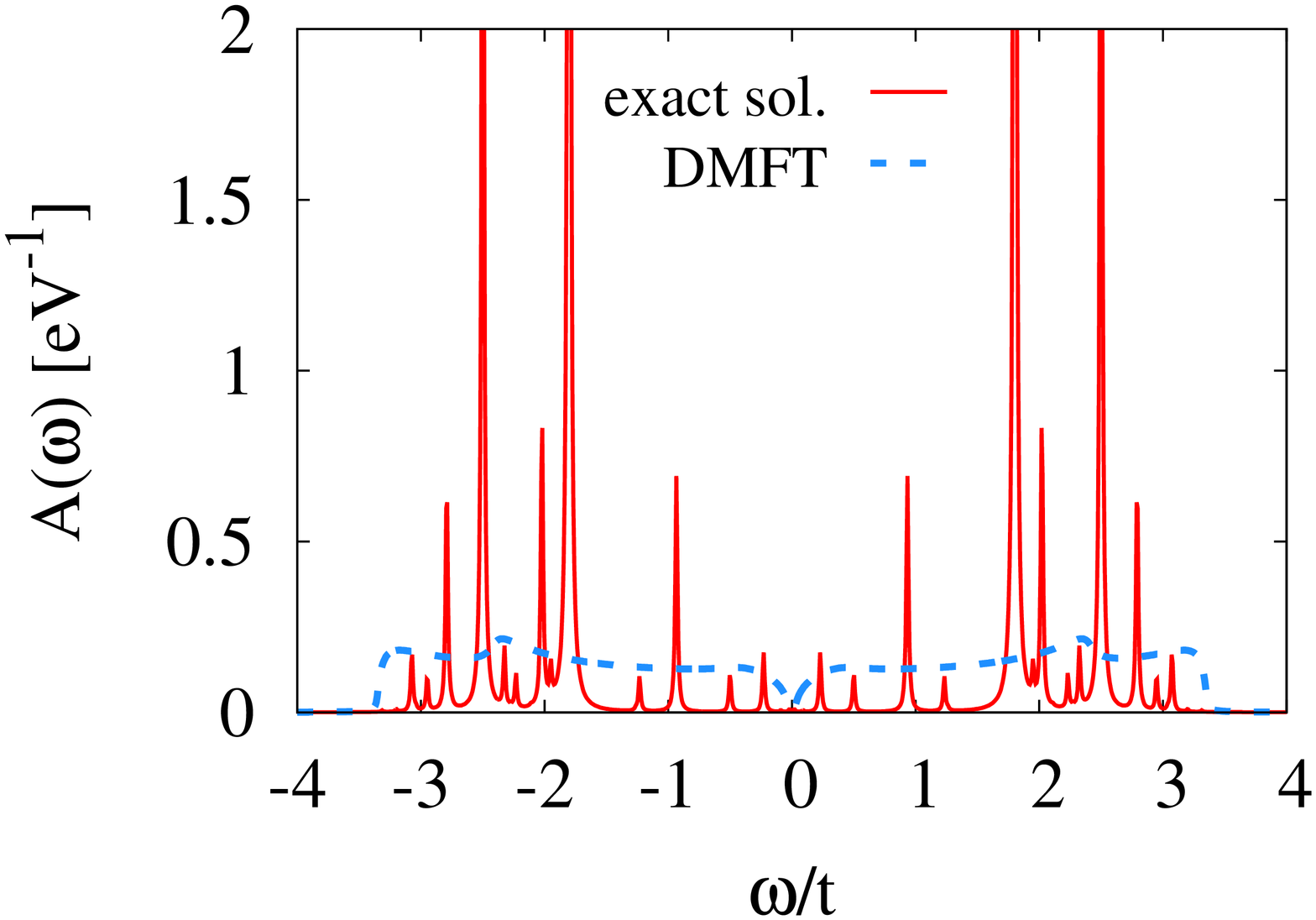}
 \includegraphics[angle=270,width=0.3\textwidth]{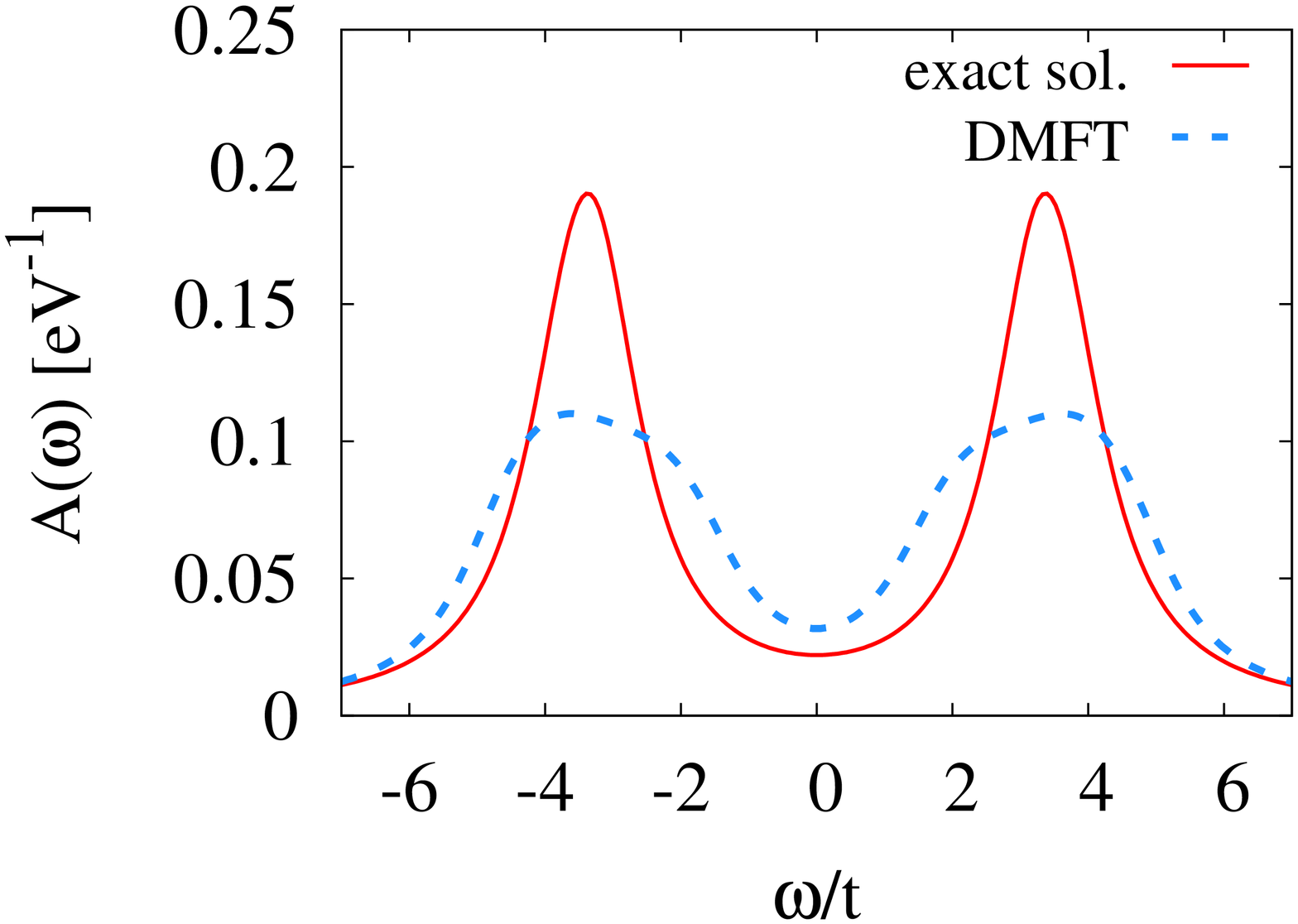}
 \includegraphics[angle=270,width=0.3\textwidth]{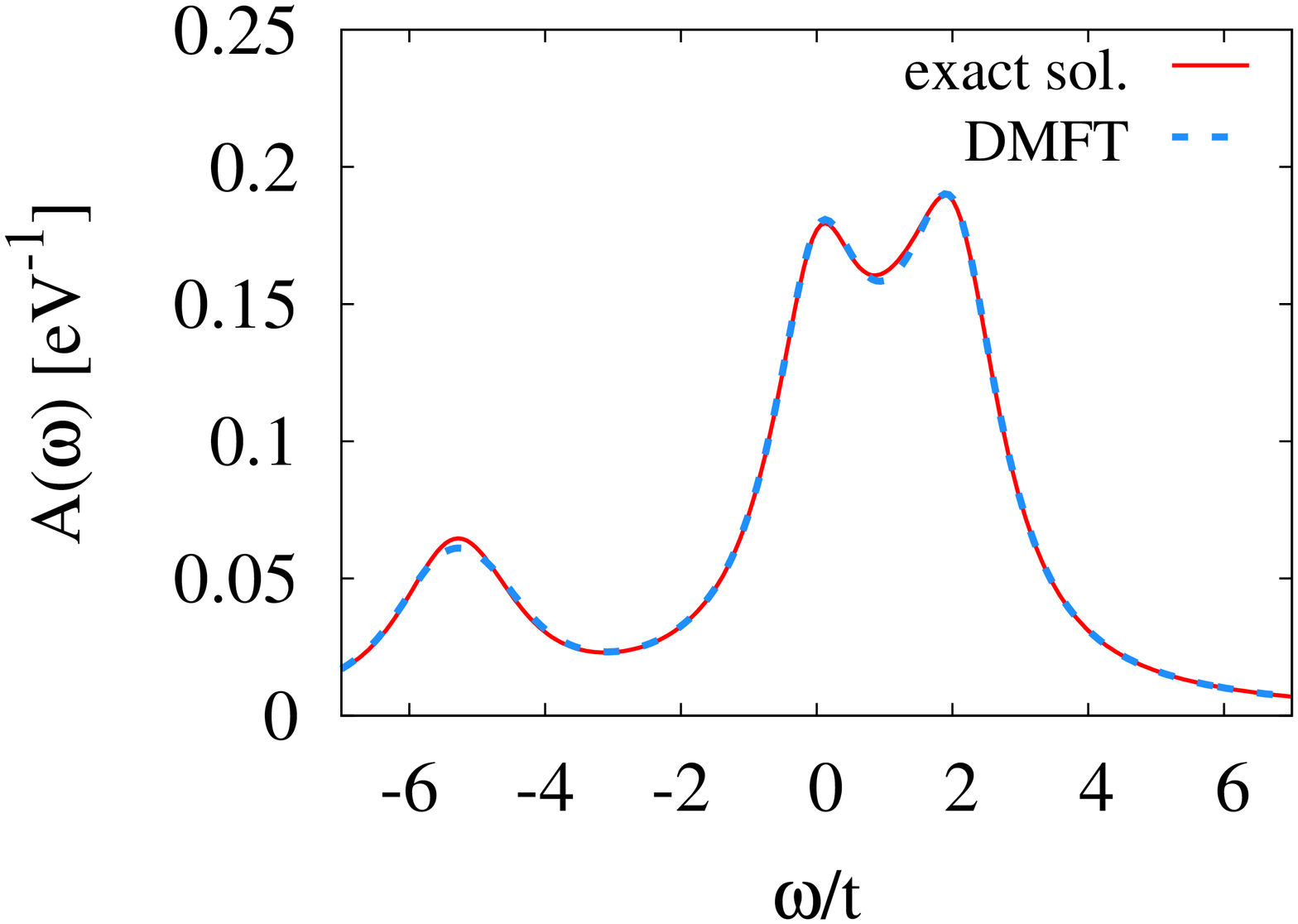}
 \caption{Representative results of the comparison of the spectral function $A(\omega )$ between the exact solution and DMFT.  
Left panel: $J=1.5t$, $V/t=0$, and $\beta t=5$ for NN~t at half-filling which implies particle hole-symmetry for  NN~t. In the low-hybridization regime the exact solution 
displays many peaks which correspond to the eigenstates of the Hamiltonian, while DMFT rather predicts broad bands. 
Middle panel: $J=3t$, $V=t$, and $\beta t=20$ for NN~t at half-filling; increasing the hybridization 
the DMFT description improves even at strong coupling. 
Right panel: $J=t$, $V=t$, and $\beta t=20$ for all~t; increasing the connectivity 
the DMFT reproduces the exact solution, even in a non-perturbative regime. 
A similar behavior is observed also for quarter-filling (in the NN~t topology) and in all other parameter regimes  investigated.}
 \label{fig:spectraCMP}
 \end{center}
\end{figure*}
	
\subsection{Comparison with DMFT} \label{sec:results:DMFT}
Let us turn to the DMFT solution, which for the specific case of the double exchange model is temperature independent
(see Eq. \eqref{ImpSolver}) and generally neglects non-local correlations. 
We hence expect DMFT to describe a system better if there is no (short-range) ordering of the spins, 
i.e., when the different spin configurations are rather equally important. 

For the NN~t system, the left panel of Fig. \ref{fig:spectraCMP} shows that 
the spectral function of the exact solution exhibits many peaks 
at energies corresponding to the $N2^N$ eigenvalues of the Hamiltonian, 
while DMFT treats these states within mean field, and therefore predicts broad bands. 
Except for this striking difference, the spectral function is qualitatively reproduced, 
and the DMFT spectrum is similar to a broadened version of the exact spectrum. 
In the presence of the hybridization $V$, the peaks get broadened anyhow. 
Hence, the finite-$V$ spectrum predicted by DMFT is more similar to the exact one, 
as can be observed in the middle panel of Fig. \ref{fig:spectraCMP}. 
In the all~t topology, where non-local correlations are washed away due to the high connectivity, 
the exact spectral function is almost exactly reproduced by DMFT, 
as shown in the right panel of Fig. \ref{fig:spectraCMP}.

\begin{figure}
 \begin{center}
 \includegraphics[angle=270,width=0.5\textwidth]{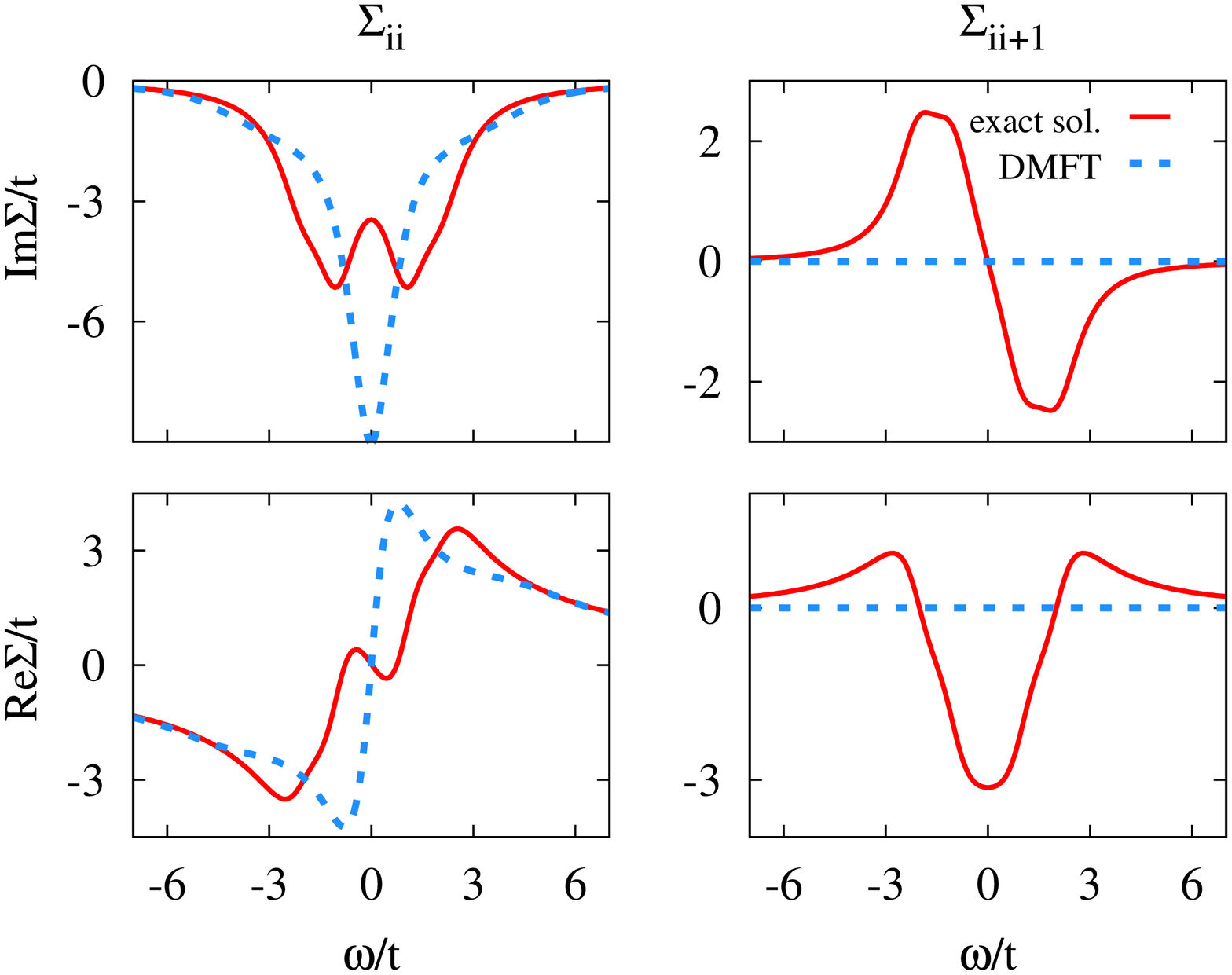}
 \caption{Local ($\Sigma_{ii}$, left panels) 
and non-local (for neighboring sites $\Sigma_{i,i+1}$, right panels) self-energy for $J\!=\!3t$, $V\!=\!t$, $\beta t\!=\!20$, and
NN~t hopping topology.} 
 \label{fig:sre_NNt}
 \end{center}
\end{figure}	
\begin{figure}
 \begin{center}
 \includegraphics[angle=270,width=0.5\textwidth]{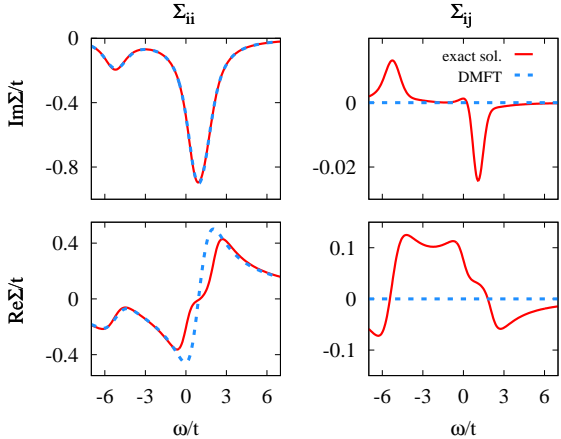}
 \caption{Local ($\Sigma_{ii}$, left panels) 
and non-local ($\Sigma_{ij}$, equivalent for each pair $i\!\neq\!j$, right panels) 
elements of the self-energies for $J\!=\!t$, $V\!=\!t$, and $\beta t\!=\!20$, 
for both all~t hopping topology.} 
 \label{fig:sre_allt} 
 \end{center}
\end{figure}

A more precise characterization of the non-local spatial correlations between neighboring localized spins 
can be achieved considering the self-energies. In Fig. \ref{fig:sre_NNt} and \ref{fig:sre_allt} 
we compare the exact and DMFT local ($\Sigma_{ii}$) and the non-local 
(between neighboring sites, $\Sigma_{i,i+1}$) self-energies corresponding to the spectral functions 
shown in the middle and right panel of Fig. \ref{fig:spectraCMP}, respectively.

In a Fermi liquid ground state, the local self-energy displays, around the Fermi energy $E_F$, a typical behavior 
$\Sigma_{ii}(0)\sim-\alpha \omega +\imath(\gamma+\delta\omega^2)$ with the coefficients $\alpha>0$ and $\gamma,\delta<0$. 
In the NN~t case, shown in Fig. \ref{fig:sre_NNt}, the DMFT local self-energy is highly non-Fermi liquid, 
displaying a large maximum (in absolute value) in the imaginary part of $\Sigma$ 
and a positive slope of the real part ($\alpha<0$). These are the fingerprints 
of the depletion at $E_F$ observed in the corresponding spectral function $A(\omega)$, 
which is also a consequence of the band gap that characterizes the non-interacting ring. 
Interestingly, the exact local self-energy behaves, in contrast, like in a Fermi liquid, 
and hence the comparison with DMFT looks poor. 
In absence of non-local correlations, the exact local self-energy would correspond to a scenario 
in which $A(\omega)$ would display a (renormalized) quasi-particle peak. 
The exact solution also predicts a pseudogap. However, this is 
generated by non-local magnetic fluctuations, e.g., by Re$\Sigma_{i,i+1}(0)$ between nearest neighbors (cf. also \cite{toschiPRB75,kataninPRB80,maierRMP77}). 
Non-local spatial correlations extend also to longer distances, 
and the corresponding elements of the self-energy (not shown) are quantitatively comparable 
to the nearest-neighbor one shown here.

In the all~t case, shown in Fig. \ref{fig:sre_allt}, the picture is qualitatively the same, 
with the fundamental difference, that the non-local contributions to the self-energy 
are sensibly smaller than the local ones, thus justifying that the DMFT spectral function well agrees with the exact one. 
Remarkably, due to the enhanced connectivity, the non-local self-energies 
remain small also at lower values of $V/t$, 
when the suppression of magnetic correlations due to the hybridization is less effective (not shown).
This shows that for high enough connectivity, or strong hybridization, 
the Hund's coupling most likely does not result in a phase with short range order, 
and  therefore can be  treated reliably within a dynamical mean field approximation.

\section{Conclusion and outlook} \label{sec:conclusion}
We have studied the double exchange model on a nanoscopic lattice. 
The advantage of this simple model is that it can be solved exactly also for finite systems. 
This allows us to systematically study and understand the nature of approximations such as DMFT. 
Hence we can address the question: Where is this approximation sufficient, 
where does it fail quantitatively or qualitatively? 
We found DMFT to be reliable in a wide range of parameters. 
In particular we investigated the role of non-local correlations which are neglected in any DMFT-like calculation. 
These non-local magnetic correlations are suppressed if, 
e.g., the nano-cluster strongly hybridizes with metallic leads, 
or if it has a large coordination number, explaining the overall good quality of the DMFT predictions. 
After our analysis we can conclude that DMFT can be a suitable tool 
to study also more complex correlated nanostructures in this parameter range. 
However, if non-local correlations are not negligible, DMFT is not reliable anymore, 
and a better approximation is needed. In this respect, it is possible to extend the method 
in the spirit of D$\Gamma$A \cite{nanoDGA,toschiPRB75,kataninPRB80,kusunoseJPSJ75,slezakJPCM21,rohringerPRB86}, 
which can account for those non-local correlations that are due to local fully irreducible two-particle vertices. 
This step is expected to improve the approximation significantly. 

\section*{Acknowledgments}
We would like to thank A.~Toschi and G.~Rohringer for useful discussions. 
We also acknowledge financial support from the Austrian Science Fund (FWF) 
through the Austria-Russia joint Project No. I610 (AV), 
I597 which is part of the DFG research unit FOR1346 (DR), 
and the Austrian Bundesministerium f\"{u}r Wissenschaft und Forschung 
and the European Union within the EU-Indian network MONAMI (KH).

%
% Non-BibTeX users please use


\begin{thebibliography}{}

%\bibitem{AM} N.~W.~Ashcroft and N.~D.~Mermin, Solid State Physics, Holt, Rinehart and Winston (1976).
\bibitem{CMR} 
R.~von~Helmolt, J.~Wecker, B.~Holzapfel, L.~Schultz, and K.~Samwer Phys. Rev. Lett. \textbf{71}, 2331 (1993);
K. Chahara, T. Ohno, M Kasai, and Y. Kozono, Appl. Phys. Lett. \textbf{63}, 1990 (1993); 
S.~Jin, T.~H.~Tiefel, M.~McCormack, R.~A.~Fastnacht, R.~Ramesh, and L.~H.~Chen, Science {\bf 264}, 413 (1994); 
P.~Schiffer, A. P. Ramirez, W. Bao, S-W. Cheong, Phys. Rev. Lett. \textbf{75}, 3336 (1995).

\bibitem{DE} C.~Zener, Phys. Rev. {\bf 82}, 403 (1951).

%\bibitem {MillisJT} 
%A.~J.~Millis, P.~B.~Littlewood, and B.~I.~Shraiman, Phys. Rev. B \textbf{74}, 5144 (1995); 
%A.~J. Millis, R.~Mueller and B.~I. Shraiman, Phys. Rev. B {\bf 54} 5405 (1996); Phys. Rev. B {\bf 54} 5405 (1996).

%\bibitem {Held} 
%K.\ Held and D.\ Vollhardt, Phys. Rev. Lett. 84, 5168 (2000); 
%A. Yamasaki, M. Feldbacher, Y.-F. Yang, O. K. Andersen and K. Held, Phys. Rev. Lett. \textbf{96}, 166401 (2006); 
%Y.-F.~Yang and K.~Held, Phys. Rev. B, {\bf 82}, 195109 (2010).

\bibitem{sarkar123104} 
T.~Sarkar, A.~K.~Raychaudhuri, and T.~Chatterji, Appl. Phys. Lett. {\bf 92}, 123104 (2008);  
Appl. Phys. Lett {\bf 92}, 123104 (2008); 
J. App. Phys. {\bf 101}, 124307 (2007); 
S. S. Rao {\it et. al.} App. Phys. Lett. {\bf 87}, 182503 (2005).

\bibitem{Das}
H. Das, G. Sangiovanni, A. Valli, K. Held, and T. Saha-Dasgupta,
Phys. Rev. Lett. {\bf 107}, 197202 (2011).

\bibitem{nanoDGA} 
A.~Valli, G.~Sangiovanni, O.~Gunnarsson, A.~Toschi, and K.~Held, Phys. Rev. Lett. {\bf 104}, 246402 (2010); 
A.~Valli, G.~Sangiovanni, A.~Toschi, and K.~Held, Phys. Rev. B {\bf 86} 115418 (2012).

\bibitem{DMFT}
W.~Metzner and D.~Vollhardt, \newblock Phys. Rev. Lett. {\bf 62}, 324 (1989);
A.~Georges and G.~Kotliar, \newblock Phys. Rev. B {\bf 45}, 6479 (1992);
A.~Georges, G.~Kotliar, W.~Krauth and M.~Rozenberg, \newblock Rev. Mod. Phys. {\bf 68}, 13 (1996).

\bibitem{Furukawa1994} 
N.~Furukawa, J. Phys. Soc. Jpn., {\bf 63}, 3214 (1994);
N.~Furukawa, {\em Physics of Manganites\/}, edited by T.~A. Kaplan and S.~D. Mahanti (Kluwer, New York, 1999).


\bibitem{NegOrl} W.~Negele and H.~Orland, \newblock {\em Quantum Many-Particle Systems\/} (Addison-Wesley, New York, 1987).

\bibitem{Florens08a} S.~Florens, Phys. Rev. Lett. {\bf 99}, 046402 (2007).

\bibitem{Potthoff99} M.~Potthoff and W.~Nolting, Phys. Rev. B {\bf 59}, 2549 (1999).
 %\bibitem[{\citenamefont{Potthoff and
 %  Nolting}(1999{\natexlab{a}})}]{Potthoff99}
 %\bibinfo{author}{\bibfnamefont{M.}~\bibnamefont{Potthoff}} \bibnamefont{and}
 %  \bibinfo{author}{\bibfnamefont{W.}~\bibnamefont{Nolting}},
 %  \bibinfo{journal}{Phys. Rev. B} \textbf{\bibinfo{volume}{59}},
 %  \bibinfo{pages}{2549} (\bibinfo{year}{1999}{\natexlab{a}}),
 %  \bibinfo{pages}{{\it ibid}} \textbf{\bibinfo{volume}{60}},
 %  \bibinfo{pages}{7834} (\bibinfo{year}{1999}{\natexlab{b}}).

\bibitem{Snoek08} M.~Snoek, I.~Titvinidze, C.~T\H{o}ke, K.~Byczuk, and W.~Hofstetter, New J. Phys. {\bf 10}, 093008 (2008).
 %\bibitem[{\citenamefont{Snoek et~al.}(2008)\citenamefont{Snoek, Titvinidze,
 % T{\"o}ke, Byczuk, and Hofstetter}}]{Snoek08}
 %\bibinfo{author}{\bibfnamefont{M.}~\bibnamefont{Snoek}},
 %  \bibinfo{author}{\bibfnamefont{I.}~\bibnamefont{Titvinidze}},
 %  \bibinfo{author}{\bibfnamefont{C.}~\bibnamefont{T{\"o}ke}},
 %  \bibinfo{author}{\bibfnamefont{K.}~\bibnamefont{Byczuk}}, \bibnamefont{and}
 %  \bibinfo{author}{\bibfnamefont{W.}~\bibnamefont{Hofstetter}},
 % \bibinfo{author}{\bibnamefont{{\sl et al.}}},
 %  \bibinfo{journal}{New J. Phys.} \textbf{\bibinfo{volume}{10}}
 %  (\bibinfo{year}{2008}).


\bibitem{toschiPRB75} 
A.~Toschi, A.~A. Katanin, and K.~Held, Phys. Rev. B \textbf{75}, 045118 (2007); 
K.~Held, A.~A.~Katanin, and A.~Toschi, Prog. Theor. Phys. Supp. \textbf{176}, 117 (2008).

\bibitem{kataninPRB80} A.~A.~Katanin, A.~Toschi, and K.~Held, Phys. Rev. B \textbf{80}, 075104 (2009).
%\bibitem{rohringerPRL107} G. Rohringer, A. Toschi, A. Katanin, and K. Held, Phys. Rev. Lett. {\bf 107}, 256402 (2011).

\bibitem{henningPRB86} S.~Henning, P.~Herrmann, and W.~Nolting, Phys. Rev. B {\bf 86}, 085101 (2012).


\bibitem{maierRMP77} T.~Maier, M.~Jarrell, T.~Pruschke, and M.~H.~Hettler, Rev. Mod. Phys., {\bf 77}, 1027 (2005).


\bibitem{kusunoseJPSJ75} H.~Kusunose, {J. Phys. Soc. Jpn.} \textbf{75}, 054713 (2006).
\bibitem{slezakJPCM21} C.~Slezak, M.~Jarrell, T.~Maier, and J.~Deisz, J. Phys.: Condens. Matter {\bf 21}, 435604 (2009).

%\bibitem{Abrik} A.~A.~Abrikosov, Methods of quantum field theory in statistical physics, Prentice-Hall (1963).

\bibitem{rohringerPRB86} The fully irreducible two-particle vertex has been recently calculated in G.~Rohringer, A.~Valli, and A.~Toschi, Phys. Rev. B {\bf 86}, 125114 (2012).



%
% and use \bibitem to create references.
%
%\bibitem{RefJ}
% Format for Journal Reference
%Author, Journal \textbf{Volume}, (year) page numbers.
% Format for books
%\bibitem{RefB}
%Author, \textit{Book title} (Publisher, place year) page numbers
% etc

\end{thebibliography}
\end{document}